\documentclass[12pt,preprint]{aastex}

\newcommand{\be}{\begin{equation}}
\newcommand{\ee}{\end{equation}}

\def\lta{\,\raise 0.3 ex\hbox{$ < $}\kern -0.75 em
 \lower 0.7 ex\hbox{$\sim$}\,}
\def\gta{\,\raise 0.3 ex\hbox{$ > $}\kern -0.75 em
 \lower 0.7 ex\hbox{$\sim$}\,} 

\newcommand{\source}{{\cal S}}
\newcommand{\uout}{u_{\rm out}}
\newcommand{\uzero}{u_{0}}
\newcommand{\rout}{R_{\rm out}}
\newcommand{\vrad}{\eta} 
\newcommand{\mdotin}{ {\dot M}_{\rm in}}
\newcommand{\mjup}{M_{\rm Jup}}
\newcommand{\taueff}{\tau_{\rm eff}} 
\newcommand{\tcool}{t_{\rm cool}}

\newcommand{\uenv}{{\cal U}_{\rm env}} 

\begin{document} 

\title{\bf General Analytic Solutions for Circumplanetary Disks\\
  during the Late Stages of Giant Planet Formation} 

\author{Fred C. Adams$^{1,2}$ and Konstantin Batygin$^3$}

\affil{$^1$Physics Department, University of Michigan, Ann Arbor, MI
  48109, USA}

\affil{$^2$Astronomy Department, University of Michigan, Ann Arbor, MI
  48109, USA}

\affil{$^3$Division of Geological and Planetary Sciences, California
  Institute of Technology, Pasadena, CA 91125, USA} 

\begin{abstract}
  Forming giant planets are accompanied by circumplanetary disks, as
  indicated by considerations of angular momentum conservation,
  observations of candidate protoplanets, and the satellite systems of
  planets in our Solar System. This paper derives surface density
  distributions for circumplanetary disks during the final stage of
  evolution when most of the mass is accreted.  This approach
  generalizes previous treatments to include the angular momentum bias
  for the infalling material, more accurate solutions for the incoming
  trajectories, corrections to the outer boundary condition of the
  circumplanetary disk, and the adjustment of newly added material as
  it becomes incorporated into the Keplerian flow of the pre-existing
  disk. These generalizations lead to smaller centrifugal radii,
  higher column density for the surrounding envelopes, and higher disk
  accretion efficiency. In addition, we explore the consequences of
  different angular distributions for the incoming material at the
  outer boundary, with the concentration of the incoming flow varying
  from polar to isotropic to equatorial. These geometric variations
  modestly affect the disk surface density, but also lead to substantial
  modification to the location in the disk where the mass accretion
  rate changes sign. This paper finds analytic solutions for the
  orbits, source functions, surface density distributions, and the
  corresponding disk temperature profiles over the expanded parameter
  space outlined above.
\end{abstract}


\section{Introduction}  
\label{sec:intro}

Within the context of the core accretion paradigm for giant planet
formation \citep{bodenheimer1986,pollack1996,hubickyj2005}, most of
the mass is assembled during the third and final stage. During this
epoch, the forming system consists of a central planetary body
surrounded by a circumplanetary disk, which is enclosed within an
infalling envelope of gas and dust. The effective sphere of influence
of this forming planet extends (approximately) out to the Hill radius
$R_H$, which marks the boundary between the planet and its parental
circumstellar disk.\footnote{For completeness we note that both the
Bondi radius $R_B=GM/v_s^2$ and the scale height $H=v_s/\Omega$ of the
background circumstellar disk can also play a role (where $v_s$ is the
sound speed). For most of the parameter space of interest here, the
Hill radius $R_H<R_B$ and provides the relevant outer boundary.} The
goal of this paper is to provide a generalized analytic treatment of
the gas dynamics during the final phases of giant planet formation
when most of the mass is gathered. In particular, we find the surface
density distribution for the circumplanetary disk. In addition to
accounting for the angular momentum budget, these disks play an
important role in determining the spectral energy distributions of
forming planets and setting the initial conditions for satellite
formation. Considerations of circumplanetary disk formation are
applicable to systems where cooling of the background gas (from the
circumstellar disk) is efficient, so that the disk and planet can
form.

Angular momentum represents an important issue during this latter
phase of formation. In approximate terms, the material entering the
Hill sphere has a rotation rate comparable to that of the mean motion
$\Omega$ of the planetary orbit. With the resulting large supply of
angular momentum, most of the incoming material cannot fall all of the
way to the planetary surface. Instead, it must collect into a
cicumplanetary disk. Through the action of viscous torques, this disk
can transfer most of the mass inward, and most of the angular momentum
outward, thereby allowing the planet to form. In steady-state, the
disk adjusts its surface density so that mass accretion attains a 
constant rate (inward) in the inner disk, with a corresponding
outward flow in the outer regions that recycles gas back into the
parental nebula. 

In order to understand the physics of mass accummulation onto forming
giant planets, we can conceptually divide the problem into regimes:
[A] The flow from the circumstellar disk into the vicinity of the
forming planet, where the boundary can be taken as the Hill sphere.
In general, the background circumstellar disk provides the net mass
infall rate $\mdotin$ into the Hill sphere, the geometric distribution
of the incoming material $f(\cos\theta_0)$, and the initial conditions
for the subsequent flow, namely, the velocity components specified on
$R_H$. [B] The transport of material from the outer boundary (at the
Hill sphere) to the system midplane where it joins onto the evolving
circumplanetary disk. Within the Hill sphere, incoming material
follows orbital trajectories where the gravitational potential is
dominated by that of the planet, which lies at the center of the
collapse flow. [C] The structure and evolution of the circumplanetary
disk, which has a mass supply provided by the inflowing material
described above. Figure \ref{fig:diagram} shows a schematic diagram of
this process. Building upon past work \citep{adams2022}, this paper
provides more realistic treatments for several aspects of the problem,
as outlined below:

\begin{figure}
\centering
\includegraphics[width=0.80\linewidth]{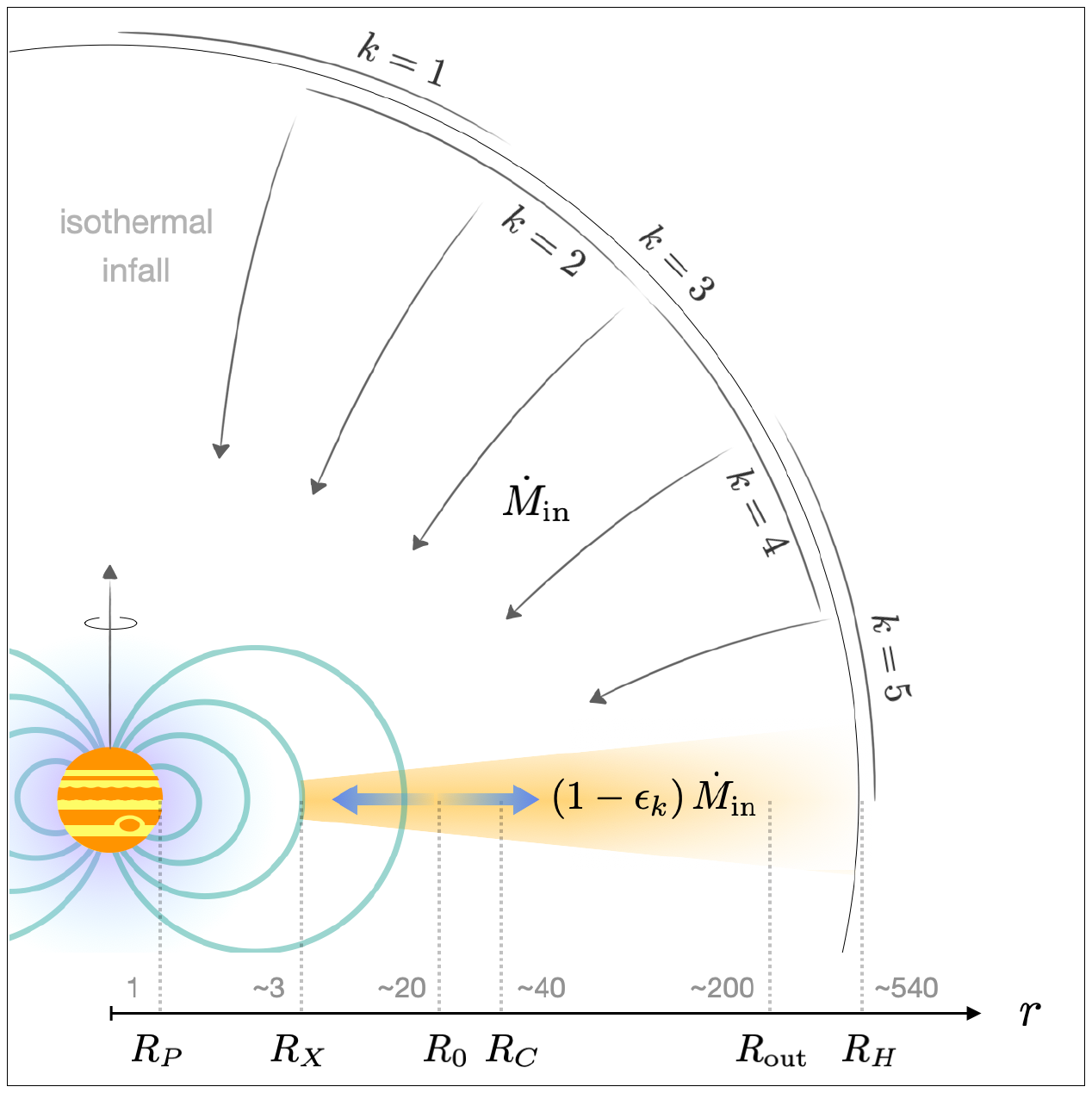}
\vskip-0.50truein
\caption{Schematic diagram of forming giant planet during the third
  stage of core accretion when most of the mass is accummulated. The
  background circumstellar disk feeds gas into the Hill sphere at a
  net rate $\mdotin$. The incoming material can have different angular
  distributions, indicated by the values of the index $k$. Infalling
  gas forms a circumplanetary disk, which in turn accretes mass onto
  the central planet with efficiency $\epsilon_k$. The length scales
  in the problem are shown along the lower horizontal axis, including
  the planet radius $R_P$, the magnetic truncation radius $R_X$ , the
  location $R_0$ where disk accretion changes sign, the centrifugal
  radius $R_C$, the outer disk radius $\rout$, and finally the Hill
  radius $R_H$. }
\label{fig:diagram}
\end{figure} 

Although the mass flowing into the Hill sphere is expected to be
rotating near the angular velocity of the local mean motion $\Omega$,
the material that enters the vicinity of the planet often has a
somewhat smaller effective rotation rate. This discepancy can be
defined in terms of an angular momentum bias factor $\lambda<1$,
defined so that the specific angular momentum in midplane is given by
$J=\lambda\Omega R_H^2$. For the simplest case of undeflected,
two-dimensional Keplerian flow, the angular momentum bias is estimated
to be $\lambda=1/4$ \citep{lissauer1991}. In general, however, the
incoming trajectories will be influenced, at least in part, by
pressure forces at large distances from the Hill sphere, and the
resulting angular momentum bias is closer to unity (see, e.g.,
\citealt{ward2010} for a more detailed discussion). In general, the
bias factor is found to be an increasing function of the ratio
$R_B/R_H$, so that it increases with the mass of the planet, and is
expected to lie in the range $\lambda=1/3-1/2$. We note that existing
calculations do not yet provide a definite prediction for this factor,
and that additional physical considerations (e.g., magnetic fields)
could affect the result.  The analytic treatment of this paper
provides solutions for any value of $\lambda$.

The angular distribution of material entering the Hill sphere has been
studied through numerical simulations of the planet formation process
(e.g., \citealt{tanigawa2012,szulagyi2017,szulagyimord,fung2019}; and
many others).  Unfortunately, these studies do not yet provide an
unambiguous specification of the expected angular dependence of the
incoming material. Some simulations indicate that the flow enters the
Hill sphere primarily along the rotational poles of the system
\citep{lambrechts2017,lambrechts2019}, whereas other simulations
suggest flow that is more concentrated toward the equatorial regions
\citep{ayliffe2009}. We also note that the angular dependence is
likely to vary with the mass of the growing planet. For example, once
the planet becomes large enough that its Hill sphere is larger than
the scale height of the circumstellar disk, the incoming flow will
become more equatorial. In light of these uncertainties, we consider a
range of possible geometries (see \citealt{taylor2024}), ranging from
flow focused along the poles with a factor $f\propto\cos^2\theta_0$,
to isotropic with $f=1$, to flow primarily along equatorial directions
with $f\propto\sin^2\theta_0$.

Next we need the orbital solutions for material entering the Hill
sphere and falling toward the central planet and its circumplanetary
disk. Past work used solutions developed earlier for star formation,
where the orbits are taken to have zero energy, zero pressure, and
where the starting radius is evaluated in the $r\to\infty$ limit
\citep{ulrich1976,cassen1981,chevalier1983, terebey1984}.  Here we
relax these assumptions (see also \citealt{mendoza2009}) and construct
orbit solutions evaluated at the finite outer boundary of the Hill
radius. We also consider a range of solutions from zero total energy
to vanishing radial velocity (at the boundary), noting that the latter
approximation is more likely to hold.

A given orbit solution, corresponding to a particular set of boundary
conditions at the Hill sphere, leads to a source term for the input of
envelope material onto the evolving circumplanetary disk.  For a given
source term, the incoming material does not, in general, have the
proper angular momentum to become part of the pre-existing
circumplanetary disk \citep{cassen1981,cassen1983}.  As a result, the
newly added material will adjust its energy, and hence is radial
location within the disk, in a manner that conserves angular momentum
and dissipates energy. This readjustment affects the evolution of the
circumplanetary disk and its steady-state surface density profile.

Finally, in order to provide a proper accounting of the angular
momentum budget for forming planets, we need to specify the outer
boundary of the circumplanetary disk. Here the default assumption
would be that the circumplanetary disk extends out to the Hill radius,
where the potential of the background circumstellar disk must
dominate. The Hill radius thus represents an upper limit to the outer
boundary. In practice, the effective outer disk boundary falls at a
smaller radius, $\rout=\xi R_H$. At sufficiently large distances from
the planet, the circumplanetary orbits become sufficiently
non-circular so that they cross, dissipate energy, and can no longer
support a coherent disk structure. Previous calculations indicate that
$\xi\approx2/5$ \citep{martin2011}.

The aforementioned numerical simulations highlight additional
complications regarding the formation of circumplanetary disks and
their host planets. Some of the material that initially enters the
Hill sphere promptly flows back out 
\citep{machida2008,lambrechts2017}. In this paper, we consider the
mass infall rate $\mdotin$ to represent the {\it net} rate at which
material enters the Hill sphere (the difference between the mass
rate inward and any outward flow). The analytic solutions found herein
hold for any value of $\mdotin$, but its value is taken as a starting
point (so that its calculation is beyond the scope of this work). 

Another important complication involves the cooling rate of the
incoming gas.  A common thread through all numerical studies of
circumplanetary hydrodynamics is the critical role of thermodynamics
in circumplanetary disk formation.  Numerical experiments (e.g.,
\citealt{szulagyi2017,fung2019,krapp2024}) consistently show that
effective radiative cooling is required to shed the heat generated by
accretion shocks and compressional heating -- only then can gas settle
into a rotationally supported disk. Without sufficient cooling, the
gas remains nearly adiabatic and forms an envelope supported by
pressure and entropy gradients instead of rotation. For example,
global 3D radiative-hydrodynamics simulations \citep{klahr2006} showed
that the forming planet was cocooned by a hot, pressure-supported
envelope in lieu of a thin disk. More recent radiative simulations
\citep{szulagyi2016} found that capping the temperature (to simulate
cooling) was necessary to obtain a disk --- implying a threshold in
cooling efficiency below which a disk cannot form. Inefficient cooling
also leads to recyling flows that channel a large fraction of gas
through the Hill sphere without necessarily leading to disk formation
\citep{lambrechts2019}. Extending these results, \cite{schulik2020}
found a sharp transition near Saturn-to-Jupiter mass scales, where
lower-mass protoplanets exhibit pressure-supported envelopes and
higher-mass planets develop disks if cooling is sufficiently effective
(see also \citealt{bailey2023}). We emphasize here that in order for
the system to accrete enough mass to produce a giant planet, the disk
structure is necessary to transfer angular momentum outward and mass
inward. In this paper, we thus consider only systems with sufficient
cooling, where a disk and a planet can form (see Appendix
\ref{sec:cooling} for additional discussion of this issue).

The analytic treatment of this paper complements existing numerical
treatments of the problem. Despite tremendous advances, present-day
simulations face significant limitations in capturing the full physics
of circumplanetary disk formation. One major challenge is resolution:
Resolving both the background disk environment and the planetary
radius itself in one simulation is difficult with current
computational resources.  Even with nested grids or mesh refinement,
the innermost flow near the planet is often smoothed or softened
(e.g., \citealt{lega2024}). Insufficient resolution can artificially
suppress small-scale structures and can affect the measured accretion
rates. For example, increasing the resolution around the planet
significantly alters the disk temperature, luminosity, and density
profile \citep{schulik2020}, indicating that some simulations are not
yet numerically converged.  The time spans simulated (typically tens
to hundreds of orbits) are short compared to those of actual planet
formation (millions of years), making it difficult to examine a true
steady state. Finally, computational costs force a trade-off between
global realism and local detail: Global disk simulations have coarse
resolution in the vicinity of the planet, whereas local `zoomed-in'
studies (e.g., \citealt{zhu2024}) may not include large-scale flow
features (e.g., the mass supply from the circumstellar disk). Due to
these limitations, current simulations cannot fully account for the
formation of a planet, disk, and envelope with all of the relevant
physics. One goal of this paper is thus to fill in some these gaps
with analytic solutions.

This paper calculates the form of the surface density for
circumplanetary disks, which play an important role in determining the
radiative signatures of forming planets. These systems are just now
becoming observable
\citep{benisty2021,bae2022,christiaens2024,cugno2024}. During planet
formation, the disk dissipates significant energy and provides much of
the luminosity. Compared to the planet itself, the radiation emanating
from the disk is emitted at longer wavelengths, which are more easily
observed.  As more observations become available, more sophisticated
theoretical models of the spectral energy distributions are warranted,
and more accurate surface density distributions are required (see,
e.g., \citealt{zhu2015,zhu2018,szulagyi2019,adams2022,taylor2025,
choksi2025}).

This paper is organized as follows. Section \ref{sec:orbit} constructs
the orbit solutions for the generalized case where the trajectories
start at the (finite) Hill radius, have arbitrary angular momentum
bias $\lambda$, and potentially non-zero starting radial velocity
(although in practice the latter correction is small). Next we find 
the corresponding density distribution, velocity field, and source
term for material falling onto the disk.  The effects of angular
momentum readjustment are derived in Section \ref{sec:adjust},
resulting in the generalized equation of motion for the evolution of
the circumplanetary disk. Steady-state solutions are found in Section
\ref{sec:steadystate}, resulting in analytic expressions for the mass
accretion rate and the surface density distribution as a function of
radial position. Solutions are found for five choices of the inflow
geometry, from polar to equatorial. These solutions, in turn,
determine the fraction of the mass entering the disk that can accrete
onto the planet, and thus determine the maximum accretion efficiency
subject to conservation of angular momentum. The paper concludes, in
Section \ref{sec:conclude}, with a summary of our results and a
discussion of their implications.

\section{Generalized Infall Collapse Solutions}  
\label{sec:orbit}

This section generalizes the standard collapse solution
\citep{ulrich1976} to include more realistic choices for the outer
boundary conditions.  For the sake of definiteness, the solutions are
fixed at the finite outer radius $R_H$ of the Hill sphere. For a given
angular momentum, the radial velocity, and hence the energy, can be
varied. The azimuthal velocity at the boundary is determined by the
angular momentum, which is set by the bias parameter $\lambda$. For
simplicity, we assume azimuthal symmetry and solid body rotation at
the Hill sphere. In other words, the angular momentum bias $\lambda$
is taken to be a constant, rather than a function of polar angle
$\theta_0$. Note that these assumptions could be relaxed in future
work. We also note that at the latest evoluationary stages, the planet
is expected to clear a wide gap in the background circumstellar disk,
so that continued accretion onto the planet occurs through streamers
(this accretion geometry will be addressed in future work).

Energy and angular momentum are conserved. In this formulation, the
angular momentum of a parcel starting at radius $r=R_H$ and angle
$\theta_0$ has the form 
\be
J = \lambda \Omega R_H^2 \sin\theta_0\,,
\ee
where $\Omega^2=GM_\ast/a^3$ determines the rotation rate at the
planet location, $\lambda$ is the angular momentum bias, and
$\theta_0$ is the starting polar angle. The energy of the parcel is
given by 
\be
E = {1\over2} v_r^2 + {J^2 \over 2r^2} - {GM\over r} \,, 
\ee
where $r$ is the radial coordinate centered on the planet and the
potential is that of a point mass $M$. The energy can be evaluated at
the outer boundary, so we have 
\be
E = E_0 = {1\over2} v_H^2 + {J^2 \over 2R_H^2} -
{GM\over R_H} \,. 
\ee
The angular momentum in this expression can be
rewritten to take the form 
\be
J^2 = \lambda^2 \Omega^2 R_H^4 \sin^2\theta_0 
= {1\over3} \lambda^2 GMR_H \sin^2\theta_0\,,
\ee
so that the energy becomes 
\be
E_0 = {1\over2} v_H^2 + {GM\over R_H} \left[
{1\over6} \lambda^2 \sin^2\theta_0 - 1 \right] \,. 
\ee
Here we define a dimensionless energy
\be
\epsilon \equiv 
E_0 \left({GM\over R_H}\right)^{-1} = {1\over2}\vrad^2
+ {1\over6} \lambda^2 \sin^2\theta_0 - 1 \,, 
\ee
where the dimensionless radial velocity $\vrad$ (defined
at the boundary) is given by 
\be
\vrad^2 = {v_H^2 R_H \over GM} \,. 
\ee
In this treatment, the starting position is taken to be the Hill
sphere ($r=R_H$) and the system is axisymmtric. The remaining initial
spatial variable is the starting polar angle $\theta_0$. The angular
momentum bias $\lambda$ determines the azimuthal speed in the orbit
plane and the value of $\vrad$ completes the specificiation of the
initial condition (see also \citealt{mendoza2009}).

The infalling parcel of gas executes an orbit in a plane that is
tilted with respect to the spherical coordinate system centered
on the planet with $\hat z$ axis along the pole. Within the tilted 
plane, the orbit equation has the usual form which can be written
\be
u = {\sin^2 \theta_0 \over 1 - e \cos\varphi} \,,
\ee
where $\varphi$ is the azimuthal angle in the
orbital plane and 
\be
u = {r \over R_C}  \qquad {\rm where} \qquad
R_C = {J^2 (\pi/2) \over GM} = {1\over3} \lambda^2 R_H\,. 
\ee
The centrifugal radius is thus smaller than in the benchmark case
(where $R_C=R_H/3$; \citealt{quillen1998}) by a factor of $\lambda^2$. 
The eccentricity of the orbit can be determined from the definitions
of energy and angular momentum of the orbit and takes the form
\be
e^2 = 1 + {2\over3} \lambda^2 \epsilon \sin^2\theta_0 \,.
\label{eccent} 
\ee
Next we use standard geometric rotations to write
\be
\cos(\varphi - \varphi_0) = {\cos\theta \over \cos\theta_0}
\qquad {\rm or} \qquad 
\varphi = \varphi_0 + \cos^{-1} \left[
  {\cos\theta \over \cos\theta_0} \right] \,. 
\ee

\subsection{Effective Initial Disk Radius}
\label{sec:diskradius} 

Now consider the trajectory that leads to the largest radius that
crosses the midplane, i.e., the effective disk radius. Note that this
radius is the initial effective radius of the disk, which will
subsequently spread outward due to the action of viscosity (see
Section \ref{sec:steadystate}).  Here we first take
$\cos\theta=\mu=0$.  The angle $\varphi$ is then given by
\be
\varphi = \varphi_0 + {\pi\over2} \,,
\ee
so that
\be
\cos\varphi = \cos(\varphi_0) \cos(\pi/2) -
\sin(\varphi_0) \sin(\pi/2) = - \sin(\varphi_0) \,. 
\ee
If we evaluate the orbit equation at the outer boundary where
$u=3/\lambda^2$ and $\varphi$ = $\varphi_0$, we can solve to find
\be
e \cos\varphi_0 = 1 - {1\over3} \lambda^2 \sin^2\theta_0 \,. 
\ee
We can use this expression to find $e\sin\varphi_0$.
Using this result in conjunction with the expression of equation
(\ref{eccent}) for the eccentricity, we find that 
\be
e^2 \sin^2\varphi_0 =
1 + {2\over3} \lambda^2 \epsilon \sin^2\theta_0 -
\left[1 - (\lambda^2/3) \sin^2\theta_0 \right]^2\,.
\ee
After inserting the expression for the energy $\epsilon$, we find 
\be
e^2 \sin^2\varphi_0 = \vrad^2 \lambda^2 \sin^2\theta_0/3 \qquad {\rm or}
\qquad e\sin\varphi_0 = \vrad \lambda \sin\theta_0 / \sqrt{3} \,. 
\ee
The orbit equation then takes the form 
\be
u = {\sin^2\theta_0 \over 1 + \vrad \lambda \sin\theta_0/\sqrt{3}} \,. 
\ee
The nominal disk radius is set by the maximum value of $u$, which
occurs for $\sin\theta_0=1$; these orbits start in the equatorial
plane and have the largest specific angular momentum. The disk
radius is thus given by 
\be
R_{\rm d} = {R_C \over 1 + \vrad\lambda/\sqrt{3}} =
{\lambda^2 R_H \over 3 + \sqrt{3} \lambda \vrad} \,.
\label{diskrad} 
\ee
The disk radius can be somewhat smaller than the centrifugal radius,
although we expect both $\vrad$ and $\lambda$ to be small, so the
correction in the denominator is modest.\footnote{Note that the above
argument takes the limit $\mu\to0$ first, and then takes the limit
$\mu_0\to0$. The order matters. If one takes the limit $\mu_0\to0$
first, then the orbit is just the standard Keplerian orbit in the
equatorial plane, and $\mu=\mu_0=0$ everywhere. One could find the
inner turning point of the orbit, but that location must be inside the
true disk radius. In the limit where $\mu_0\ll1$ (not still non-zero)
the orbit crosses the equatorial plane before it reaches its inner
turning point.}

\subsection{Limit of Zero Initial Radial Velocity}
\label{sec:zerovlimit} 

The starting velocity at the Hill sphere is expected to be relatively
small. For example, if we take $v_H$ to be the sound speed, then
$\vrad^2 \sim 0.07$ for Jovian planets forming near the ice line. The
kinetic energy from the radial velocity is thus only $\sim3.5\%$ of
the gravitational potential energy. This small value for the starting
velocity motivates us to consider the limiting case where
$\vrad\to0$. In this limit, the energy is given by
\be
\epsilon = {1\over6} \lambda^2 \sin^2\theta_0 - 1 \,,
\ee
and the eccentricity is given by
\be
e^2 = 1 + {2\over3} \lambda^2 \sin^2\theta_0 \left[
  {1\over6} \lambda^2 \sin^2\theta_0 - 1 \right] =
\left[ 1 - {1\over3} \lambda^2 \sin^2\theta_0 \right]^2 \,,
\ee
so that
\be
e = 1 - {1\over3} \lambda^2 \sin^2\theta_0 \,. 
\ee
If we evaluate the orbit equation at the boundary
to specify the starting angle $\varphi_0$, we find
\be
1 - \left[ 1 - {1\over3} \lambda^2 \sin^2\theta_0 \right]
\cos\varphi_0 = {1\over3} \lambda^2 \sin^2\theta_0
\qquad \Rightarrow \qquad
\cos\varphi_0 = 1 \,, 
\ee
or, equivalently, $\varphi_0=0$. The general form of the
orbit equation thus becomes
\be
1 - \left[ 1 - {1\over3} \lambda^2 (1-\mu_0^2) \right]
{\mu\over\mu_0} = {1 \over u} (1-\mu_0^2) \,.
\label{zerovelorbit} 
\ee
To find the disk radius, we can evaluate this expression at
$\mu=0$, and then take the limit $\mu_0\to0$  to find
\be
u = 1 \qquad \Rightarrow \qquad R_{\rm d} = R_C \,.
\label{diskradzero} 
\ee
In this limit, we can also find the radial location $u_d$ where
incoming trajectories strike the disk. Evaluating the orbit
equation in the limit $\mu\to0$, we find the standard result,
\be
u_d (\mu_0) = 1 - \mu_0^2 \,,
\label{uvangle} 
\ee
i.e., the same as before in the Ulrich limit
(but not for the case with $\eta\ne0$).

The solutions of this subsection are carried out in the limit
$\eta\to0$. To quantify the severity of this assumption, one can  
consider how nonzero radial velocity affects the nominal disk size.
Suppose that the radial velocity at the Hill sphere is given by the
local Keplerian shear so that $v_H = \Omega R_H$ and $\eta^2=1/3$.
In this case, the nominal disk radius is no longer given by equation
(\ref{diskradzero}), but rather by equation (\ref{diskrad}) with
$\eta\ne0$. The relative difference in disk radius can be written 
\be
{\Delta R_{\rm d} \over R_{\rm d}} =
{\lambda\eta \over \sqrt{3} + \lambda \eta} \rightarrow
{1\over7} \,, 
\ee
where the final equality holds for $\eta=1/\sqrt{3}$ and
for an angular momentum bias $\lambda=1/2$. The correction due to
nonzero starting radial velocities is thus modest.  For completeness,
note that infall solutions with $\eta\ne0$ can be readily constructed,
although the expressions become much more complicated (Adams \&
Batygin -- unpublished). We leave a more general treatment of this
problem for future work (see also \citealt{mendoza2009}). 

\subsection{Velocity Fields}
\label{sec:vfield} 

We can find the velocity fields by applying conservation of angular
momentum and energy. The azimuthal velocity is given by conservation
of the $\hat z$ compontent of angular momentum,
\be
v_\phi = {J_z \over r\sin\theta} \qquad {\rm where} \qquad
J_z = \lambda \Omega R_H^2 \sin^2\theta_0 =
\sqrt{GMR_C} \sin^2\theta_0 \,,
\ee
so we find
\be
v_\phi = \left({GM\over R_C}\right)^{1/2} {R_C\over r}
{\sin^2\theta_0 \over \sin\theta} \,.
\label{vphi} 
\ee
Similarly, the velocity component $v_\theta$ is determined
by requiring conservation of total angular momentum, i.e.,
\be
r^2 (v_\theta^2 + v_\phi^2) = J_T^2 = GMR_C \sin^2\theta_0 \,.
\ee
Using the previous result, we find
\be
v_\theta = \left({GM\over R_C}\right)^{1/2}
{R_C\sin\theta_0\over r\sin\theta}
\left[\cos^2\theta_0 - \cos^2\theta\right]^{1/2} \,.
\label{vtheta} 
\ee
Note that equation (\ref{vphi}) and (\ref{vtheta}) were derived
using conservation of angular momentum, and we did not need to
invoke conservation of energy. As a result, these expressions
are valid for all choices of the orbital energy.

For the radial velocity component, we need to specify the orbital
energy or equivalently the value of $\vrad$. The energy is given by 
\be
E_0 = {1 \over 2} v_H^2 + {GM\over R_H} \left[ {1\over6}
  \lambda^2 \sin^2\theta_0 -  1 \right] = {GM\over 2R_H} \left[
  \vrad^2 + {1\over3} \lambda^2 \sin^2\theta_0 -  2 \right] \,.
\ee
The radial velocity thus takes the form 
\be
v_r = \left( {GM\over R_C} \right)^{1/2}
\left\{ {1\over3}\lambda^2 
\left[ \vrad^2 + {1\over3} \lambda^2 \sin^2\theta_0
-  2 \right] - \zeta^2 \sin^2\theta_0 + 2 \zeta \right\}^{1/2} \,.
\label{vradial} 
\ee
In the limit of zero initial velocity, the expression becomes
\be
v_r = \left( {GM\over R_C} \right)^{1/2}
\left\{ {1\over3}\lambda^2 
\left[ {1\over3} \lambda^2 \sin^2\theta_0
-  2 \right] - \zeta^2 \sin^2\theta_0 + 2 \zeta \right\}^{1/2} \,.
\label{vradvzero} 
\ee

\subsection{Source Term for Infall onto the Disk}
\label{sec:source} 

In order to specify the flow of material onto the circumplanetary
disk, we need to evalute the source term
\be
\source = \left[2\rho v_\theta\right]_{\mu=0} \,. 
\ee
We thus need to evalute the density and the velocity component
$v_\theta$ as the parcels of gas cross the midplane. The density
is given by conservation of mass along streamtubes, i.e.,
\be
\rho = {\mdotin f(u) \over 4\pi r^2 |v_r|} {d\mu_0\over d\mu} \,,
\ee
where all quantities are evaluated at $\mu=0$. The total net rate of
mass infall onto the disk is $\mdotin$, and the function $f(u)$
accounts for the angular dependence. More specifically, $f(u)$ = 1 for
isotropic infall, but takes on non-constant forms for infall patterns
with angular dependence (see Section \ref{sec:steadystate}). 

The polar velocity $v_\theta$ is given by equation (\ref{vtheta}),
which can be evaluated to obtain
\be
v_\theta\Big|_{\mu=0} =
\left({GM\over R_C}\right)^{1/2}
{R_C\over r} \sin\theta_0 \cos\theta_0 \,,
\ee
where the orbit equation determines $\theta_0$. In the limit of zero
starting velocity, the orbit equation can be evaluated at the midplane
$\mu=0$ (see equation [\ref{uvangle}]) to obtain $u=(1-\mu_0^2)$ or
$\mu_0^2 = 1-u$. The derivative $d\mu_0/d\mu$ thus takes the form
\be
{d\mu_0\over d\mu} =
{u \over 2\mu_0^2} \left[1 - {1\over3}\lambda^2 (1-\mu_0^2)\right]
= {u \over 2} {(1 - \lambda^2 u/3) \over (1-u)} \,.
\ee
The polar velocity becomes
\be
v_\theta\Big|_{\mu=0} =
\left({GM\over R_C}\right)^{1/2}
{R_C\over r} \sqrt{u} \sqrt{1-u} =
\left({GM\over r}\right)^{1/2}\sqrt{1-u} \,,
\ee
and the radial velocity becomes 
\be
v_r = \left( {GM\over R_C} \right)^{1/2}
\left\{ {1\over3}\lambda^2 
\left[ {1\over3} \lambda^2 u 
-  2 \right] + \zeta \right\}^{1/2} 
= \left( {GM\over r} \right)^{1/2} (1-\lambda^2 u/3) \,. 
\ee
After putting the pieces together, we can write the
source term in the form
\be
\source = 2 { \mdotin f(u) \over 4\pi r^2}
{\sqrt{1-u}\over 1-\lambda^2 u/3} \left({u\over2}\right)
{(1-\lambda^2u/3)\over(1-u)} =
{ \mdotin \over 4\pi r^2}
{u f(u) \over \sqrt{1-u}} \,.
\label{source} 
\ee

\section{Disk Surface Density with Angular Momentum Adjustment} 
\label{sec:adjust}

The continuity equation for the surface density of the disk
(e.g., \citealt{shu1990}) can be written in the form 
\be
{\partial\Sigma\over\partial t} = {1 \over 2\pi r}
{\partial {\dot M} \over \partial r}  +
\left[ 2 \rho v_\theta \right]_{\mu=0} \,, 
\label{continuity} 
\ee
where the final term is the source due to infall from the
protoplaneary envelope onto the disk and is given by equation
(\ref{source}).  In this formulation, the quantity ${\dot M}$ is the
mass accretion rate through the disk (as a function of $r$ in the
disk) and $\mdotin$ is the mass accretion rate onto the entire disk
(from the envelope).

In addition to convervation of mass (\ref{continuity}), angular
momentum must also be conserved. If we consider the increment of
angular momentum in a box surrounding an annulus of the disk, we have 
\be
d J_T = 2\pi r \Sigma \Gamma dr \qquad {\rm where} \qquad
\Gamma = \sqrt{G M_0 r} \,,
\ee
where the specific angular momentum $\Gamma$ of the disk material is
assumed to be Keplerian with a source mass $M_0$. The time rate of
change of the angular momentum is then given by 
\be
{\partial\over\partial t} \left[2\pi r\Sigma\Gamma\right] =
2\pi r \source \gamma_{\rm in} + {\partial\over\partial r}
\left[ {\dot M} \Gamma \right] +
{\partial {\cal T} \over\partial r} \,,
\ee
where $\gamma_{\rm in}$ is the specific angular momentum of the
incoming material, and where the torque ${\cal T}$ due to disk
viscosity $\nu$ \citep{shu1990,hartmann2009} is given by  
\be
{\cal T} = (2\pi r) \Sigma \nu r^2 {d\Omega\over dr} \,.
\ee
The specific angular momentum of the incoming material is given by
\be
\gamma_{\rm in} = \left[r v_\phi \right]_{\mu=0} =
\sqrt{G M_0 r} \sqrt{u} = \sqrt{G M_0 R_C}\, u \,. 
\ee
After some rearrangement, we can write the continuity equation
(multiplied by $\Gamma$) and the conservation of angular momentum
equation in the forms: 
\be
2\pi r \Gamma {\partial\Sigma\over\partial t} = \Gamma 
{\partial {\dot M} \over \partial r}  + 2\pi r \source \Gamma \,,
\ee
and
\be
2\pi r {\partial\over\partial t} \left[\Sigma\Gamma\right] =
2\pi r \source \gamma_{\rm in} + {\partial\over\partial r}
\left[ {\dot M} \Gamma \right] +
{\partial {\cal T} \over\partial r} \,. 
\ee
These two equations can be combined and then solved 
for the mass accretion rate ${\dot M}$ to find
\be
{\dot M} = 4\pi r^2 \Sigma { \mdotin \over 2 M_0}
- 4\pi r^2 \source (\sqrt{u}-1) - {2r \over \Gamma} 
{\partial {\cal T} \over\partial r}  \,. 
\ee
Now take the derivative with respect to the radial coordinate
$r$, and divide by $2\pi r$ to get 
\be
{1 \over 2\pi r} {\partial{\dot M}\over\partial r} = 
{\mdotin\over M_0}
\left[r {\partial \Sigma \over \partial r} + 2 \Sigma \right]
+ {3\over r} {\partial\over\partial r} \left( \sqrt{r} 
{\partial\over\partial r}\left[ \sqrt{r} \nu \Sigma\right] \right)
+ \source \left( {2 - \sqrt{u} - 2u \over 1 + \sqrt{u}} +
(1-\sqrt{u}) {2u\over f} {df\over du} \right) 
\ee
Using this result in equation (\ref{continuity}) gives us the
equation of motion for the evolution of the surface density
of the disk 
\be
{\partial\Sigma\over\partial t} = {\mdotin\over M_0}
\left[r {\partial \Sigma \over \partial r} + 2 \Sigma \right]
+ {3\over r} {\partial\over\partial r} \left( \sqrt{r} 
{\partial\over\partial r}\left[ \sqrt{r} \nu \Sigma\right] \right)
+ \source \left( {3 - 2u \over 1 + \sqrt{u}} + 
(1-\sqrt{u}) {2u\over f} {df\over du} \right) \,.
\ee
The time derivative operator corresponds to the case where $r$ is held
constant. In order to construct the solution for the case where $u$ is
held constant, we need another term in the time derivative to take
into account the time variation of the centrifugal radius, i.e.,
\be
{\partial\Sigma\over\partial t}\Bigg|_r =
{\partial\Sigma\over\partial t}\Bigg|_u +
{\partial\Sigma\over\partial u} {\partial u\over\partial t}
= {\partial\Sigma\over\partial t}\Bigg|_u - 
u {\partial\Sigma\over\partial u} { {\dot R}_C\over R_C} \,.
\ee
We can now write the partial differential equation for 
$\Sigma(u,t)$ in the form 
\be
{\partial\Sigma\over\partial t} = {\mdotin\over M_0} 2 \Sigma 
+ u{\partial\Sigma\over\partial u}
\left[{\mdotin\over M_0} + {{\dot R}_C \over R_C} \right]
+ {3 \over R_C^2}{1\over u}
{\partial\over\partial u} \left( \sqrt{u} 
{\partial\over\partial u}\left[ \sqrt{u} \nu \Sigma\right] \right)
\ee
$$
+ { \mdotin \over 4\pi R_C^2}
{f (1-\sqrt{u}) \over u \sqrt{1-u}}
\left( {3 - 2u \over 1 - u} +
{2u\over f} {df\over du} \right)  \,.
$$

To complete the specification of the problem, we need to enforce
boundary conditions. In the limit of small radius $r\to0$, the mass
accretion rate must approach a constant value (in the inward
direction). The inner boundary should be enforced at the inner
truncation radius, which is set by the magnetic field strength on the
planetary surface \citep{ghosh1978,blandford1982}. In practice,
however, we can enforce the inner boundary in the limit $r\to0$. At
the outer disk edge, the mass accretion is outward, and must have the
value required to remove essentially all of the incoming angular
momentum from the system.  We thus assume that the angular momentum
carried by the planetary rotation is negligible compared to the total
angular momentum entering the planet/disk system.\footnote{Note that a
rapidly spinning planet has angular momentum of order $J_{\rm p}\sim$
$M\sqrt{GMR_{\rm p}}$ and the angular momentum of that same mass $M$
has a value $J_{\rm in}\sim M\sqrt{GMR_C}$ when it enters the Hill
sphere. Since $R_C\gg R_{\rm P}$, the planet cannot absorb all of the
incoming angular momentum, which must subsequently be transfered
outwards by the disk.}

\section{Steady-State Solutions}
\label{sec:steadystate}

In this section we consider steady state solutions for the surface
density of the circumplanetary disk.  These solutions are valid in
the limit where the time scale for disk evolution due to viscosity
$t_\nu\sim R_C^2/\nu$ is shorter than the evolution time
$t_{\rm in}\sim M/\mdotin$. This ordering holds as long as the disk
viscosity parameter $\alpha\gta10^{-4}$ \citep{adams2022}, where this
requirement is likely to be realized (see \citealt{lesur2023} for a
recent review). In this limit, the planet mass will be large compared
to the mass of the circumplanetary disk. As a result, we can use a
single mass scale $M$ for the planet mass, the source mass for
Keplerian orbits in the disk, and the mass of the point potential that
determines the incoming trajectories.

For the case of interest where the infall has angular dependence, we
consider models where the incoming flow at the Hill sphere depends on
initial polar angle according to $f_1\propto\cos^2\theta_0$,
$f_2\propto\cos\theta_0$, $f_3=1$, $f_4\propto\sin\theta_0$, and
$f_5\propto\sin^2\theta_0$ \citep{taylor2024}. The corresponding disk
input function has additional dependence on the radial coordinate. In
terms of the variable $u=r/R_C$, the resulting functions $f_k(u)$ are
given by 
\be
f_1 = 3(1-u)\,, \qquad f_2=2(1-u)^{1/2}\,, \qquad f_3=1\,, \qquad
f_4={4\over\pi} \sqrt{u}\,, \qquad f_5={3\over2}u\,.
\label{fufun} 
\ee
For these infall geometries, the amount of incoming angular
momentum is given by the dimensionless factors 
\be
{\cal F}_k \equiv 
{ [{\dot J}_{\rm in}] \over \mdotin \sqrt{GMR_C}} =
2/5, \quad 1/2, \quad 2/3,
\quad 3/4, \quad {\rm and} \quad 4/5\,,
\label{jfive} 
\ee
where the sequence goes from more polar (with index $k=1$) to
isotropic $(k=3)$ to more equatorial $(k=5)$.

Here we apply the outer boundary condition at $\uout$, where the
required outgoing accretion rate is determined by the fractions from
equation (\ref{jfive}). As a result, the efficiency of mass accretion
increases as the incoming flow becomes more concentrated along the
poles. The dimensionless outer boundary is given by $\uout$
= $\rout/R_C$ = $3\xi/\lambda^2$, where we expect $\xi\approx2/5$
and $\lambda$ = 1/3 -- 1/2. As a result, we take $\uout=5$ as a
benchmark value (although the expressions derived below hold for
all values).  

In steady state, to leading order, the equation of motion
for the disk takes the form 
\be
\source \left[{3-2u \over 1+\sqrt{u}} + (1-\sqrt{u}){2u\over f}
{df\over du} \right] + {3 \over R_C^2}{1\over u} 
{\partial\over\partial u} \left( \sqrt{u} 
{\partial\over\partial u}\left[ \sqrt{u} \nu \Sigma\right] \right)
= 0 \,,
\label{steady} 
\ee
where we drop the subscript on the function $f(u)$. If we define
a dimensionless mass accretion rate $\gamma$ according to 
\be
\gamma = {6\pi \over \mdotin} \left( \sqrt{u}
{\partial\over\partial u}\left[ \sqrt{u} \nu \Sigma\right]\right)\,,
\ee
the equation of motion takes the form
\be
{\partial \gamma \over \partial u} = - {f(u) \over 2} 
\left[{(1-\sqrt{u}) (3-2u) \over (1-u)^{3/2}} +
{(1-\sqrt{u})\over\sqrt{1-u}} {2u\over f} {df\over du} \right] \,.
\ee
For the five cases of interest, right hand side of the equation
depends on $f(u)$, as given above, and the corresponding derivatives
\be
{2u\over f} {df\over du} = -{2u\over1-u} 
\,, \qquad -{u\over1-u}\,, \qquad 0\,, \qquad
1\,, \qquad {\rm and} \qquad 2\,. 
\ee

\subsection{Polar Flow}
\label{sec:polar}

For polar flow, where $\mdotin\propto\cos^2\theta_0$
and index $k=1$, the equation of motion becomes   
\be
{\partial\gamma\over\partial u} = - {3\over2} 
{(1-\sqrt{u}) \over (1-u)^{1/2}} (3-4u) \,, 
\ee
which can be integrated to obtain 
\be
\gamma = \gamma_{\rm in} - 1 + \sqrt{1-u} \left(1 - 4u + 3u^{3/2}\right)
\Theta(1-u) \,,
\ee
where $\Theta(x)$ is the Heaviside step function \citep{abrasteg}.
In the limit $u\to0$, the dimensionless mass accretion rate becomes
$\gamma_{\rm in}$.  Beyond $u>1$, the source term vanishes, and the
accretion rate $\gamma$ becomes constant with value given by 
\be
\gamma = \gamma_{\rm out} = \gamma_{\rm in} - 1 \,. 
\ee
In order to conserve angular momentum, so that the outward flow
of angular momentum at the outer boundary accounts for all of
the incoming angular momentum, one needs 
\be
\gamma_{\rm out} \sqrt{\uout} = - {2\over5} \qquad \Rightarrow
\qquad \gamma_{\rm in} - 1 = - {2\over5\sqrt{\uout}} \,.
\ee
We can integrate the mass accretion rate to find the
surface density profile:
\be
\sqrt{u} \nu \Sigma = { \mdotin \over 6\pi} 
\int_0^u \gamma(u) u^{-1/2} du \,,
\ee
which gives us
\be
\Sigma = { \mdotin \over 6\pi\nu_C} u^{-3/2}
\left\{ {4\over5} - {4 \sqrt{u} \over 5\sqrt{\uout}} 
- {2\over5} \sqrt{1-u} \, (1-\sqrt{u})^2
(2 - \sqrt{u} - 3u) \Theta(1-u) \right\} \,. 
\ee

\subsection{Quasipolar Flow}
\label{sec:quasipolar}

For the case where $\mdotin\propto\cos\theta_0$
and index $k=2$, the equation of motion becomes  
\be
{\partial\gamma\over\partial u} = - 3 (1-\sqrt{u}) \,,
\ee
which can be integrated to obtain 
\be
\gamma = \gamma_{\rm in} - 1 + (1-\sqrt{u})^2 (1 + 2\sqrt{u})
\Theta(1-u) \,.
\ee
In the limit $u\to0$, the dimensionless mass accretion is thus
$\gamma_{\rm in}$. Beyond $u>1$, $\gamma$ becomes constant, and we apply
the outer boundary condition as before. After integrating a second
time, the surface density takes the form 
\be
\Sigma = { \mdotin \over 6\pi\nu_C} u^{-3/2}
\left\{ 1 - \sqrt{u/\uout} + 
(2 \sqrt{u} - 2 u^{3/2} + u^2 - 1) \Theta(1-u) 
\right\} \,. 
\ee

\subsection{Isotropic Flow}
\label{sec:isotropic}

For the case of isotropic flow, the function $f=1$,
the index $k=3$, and the equation of motion becomes 
\be
{\partial \gamma \over \partial u} = - {1\over 2} 
{(1-\sqrt{u}) (3-2u) \over (1-u)^{3/2}} \,,
\ee
which can be integrated to find
\be
\gamma = \gamma_{\rm in} - 1 + \sqrt{1-u}
{1 + \sqrt{u} - u \over 1 + \sqrt{u}} \Theta(1-u) \,. 
\ee
Integrating again yields the surface density
\be
\Sigma = {\mdotin \over 9\pi\nu_C} u^{-3/2}
\left[2 - 2 \sqrt{u\over\uout} - \sqrt{1-u}
(2 - 3\sqrt{u} + u) \Theta(1-u) \right] \,. 
\ee

\subsection{Quasiequatorial Flow}
\label{sec:quasiequator}

For the case where $\mdotin\propto\sin\theta_0$ and index
$k=4$, the equation of motion takes the form 
\be
{\partial \gamma \over \partial u} = - {2 \over \pi}
\sqrt{u} \left[{(1-\sqrt{u}) (3-2u) \over (1-u)^{3/2}} +
{(1-\sqrt{u})\over\sqrt{1-u}} \right] = - {2\over\pi}
\sqrt{u} {(1-\sqrt{u}) (4-3u) \over (1-u)^{3/2}} \,.
\ee
After integrating we find
\be
\gamma = \gamma_{\rm in}-1  + {2\over\pi} \left\{ {(1+\sqrt{u}-2u)
(u-u^2)^{1/2} \over 1 + \sqrt{u}} + {\pi\over2} - 
{\rm ArcSin} \sqrt{u} \right\} \Theta(1-u) \,. 
\ee
Applying the boundary conditions and integrating allows us
to write the surface density in the form 
\be
\Sigma = { \mdotin \over 6\pi\nu_C} u^{-3/2} \Biggl[
{3\over2} (1 - \sqrt{u/\uout}) + \Biggl\{ 2 \sqrt{u} 
+ {1\over\pi} \Biggl[ \sqrt{1 - u} 
  (- 3 \sqrt{u} + 4 u - 2  u^{3/2}) 
\ee
$$
- 3{\rm ArcCos}\sqrt{u} 
- 4 \sqrt{u} {\rm ArcSin}\sqrt{u} 
\Biggr] \Biggr\} \Theta(1-u) \Biggr] 
$$

\subsection{Equatorial Flow}
\label{sec:equatorial} 

For this case, $\mdotin\propto\sin^2\theta_0$ and the index
$k=5$, so that the equation of motion becomes  
\be
{\partial \gamma \over \partial u} = - {3u\over4}
\left[ {(1-\sqrt{u}) (5-4u) \over (1-u)^{3/2}} \right] \,,
\ee
which integrates to the form
\be
\gamma = \gamma_{\rm in} - 1 +
{\sqrt{1-u} (2 + 2 \sqrt{u} + u + u^{3/2} - 3u^2)
\over 2 (1+\sqrt{u})} \Theta(1-u) \,. 
\ee
After integrating and applying the boundary conditionsm,
the surface density takes the form 
\be
\Sigma = { \mdotin \over 30\pi \nu_C} u^{-3/2} 
\left\{8 - 8 \sqrt{{u\over\uout}} + \sqrt{1 - u}
\left(-8 + 10 \sqrt{u} - 4 u + 5 u^{3/2} - 3 u^2\right)
\Theta(1-u) \right\}\,. 
\ee

\subsection{Summary}
\label{sec:disksummary}

All of the surface density profiles found in the previous
subsections can be written in the general form 
\be
\Sigma(r) = { {\dot M} \over 6\pi \nu} s_k(u) = 
{ {\dot M} \over 6\pi \nu_C} u^{-1} s_k(u) \,,
\label{gensigma} 
\ee
where the functions $s_k(u)$ depend on the flow geometry.
It is useful to collect the results for the reduced
surface density profiles $s_k(u)$ for the five cases: 
\be
s_1(u) = u^{-1/2}
\left\{ {4\over5} - {4 \sqrt{u} \over 5\sqrt{\uout}} 
- {2\over5} \sqrt{1-u} \, (1-\sqrt{u})^2
(2 - \sqrt{u} - 3u) \Theta(1-u) \right\} \,,
\label{surf1} 
\ee
\be
s_2(u) = u^{-1/2}
\left\{ 1 - {\sqrt{u} \over \sqrt{\uout}} + 
(2 \sqrt{u} - 2 u^{3/2} + u^2 - 1) \Theta(1-u) 
\right\} \,,
\label{surf2} 
\ee
\be
s_3(u) = u^{-1/2} {2\over3} 
\left\{ 2 - 2 {\sqrt{u}\over\sqrt{\uout}}  - 
\sqrt{1-u} (2 - 3\sqrt{u} + u) \Theta(1-u) \right\}\,,
\label{surf3} 
\ee
$$
s_4(u)  = u^{-1/2} \Biggl\{
{3\over2} - {3\over2}{\sqrt{u}\over\sqrt{\uout}} + \Biggl\{ 2 \sqrt{u} 
+ {1\over\pi} \Biggl[ \sqrt{1 - u} 
  (- 3 \sqrt{u} + 4 u - 2  u^{3/2}) 
$$
\be
- 3{\rm ArcCos}\sqrt{u} 
- 4 \sqrt{u} {\rm ArcSin}\sqrt{u} 
\Biggr] \Biggr\} \Theta(1-u) \Biggr\} \,,
\label{surf4} 
\ee
and finally 
\be
s_5(u) = u^{-1/2} {1\over5} 
\left\{8 - 8 {\sqrt{u}\over\sqrt{\uout}} + \sqrt{1 - u}
\left(-8 + 10 \sqrt{u} - 4 u + 5 u^{3/2} - 3 u^2\right)
\Theta(1-u) \right\}\,.
\label{surf5} 
\ee

\begin{figure}
\centering
\includegraphics[width=0.80\linewidth]{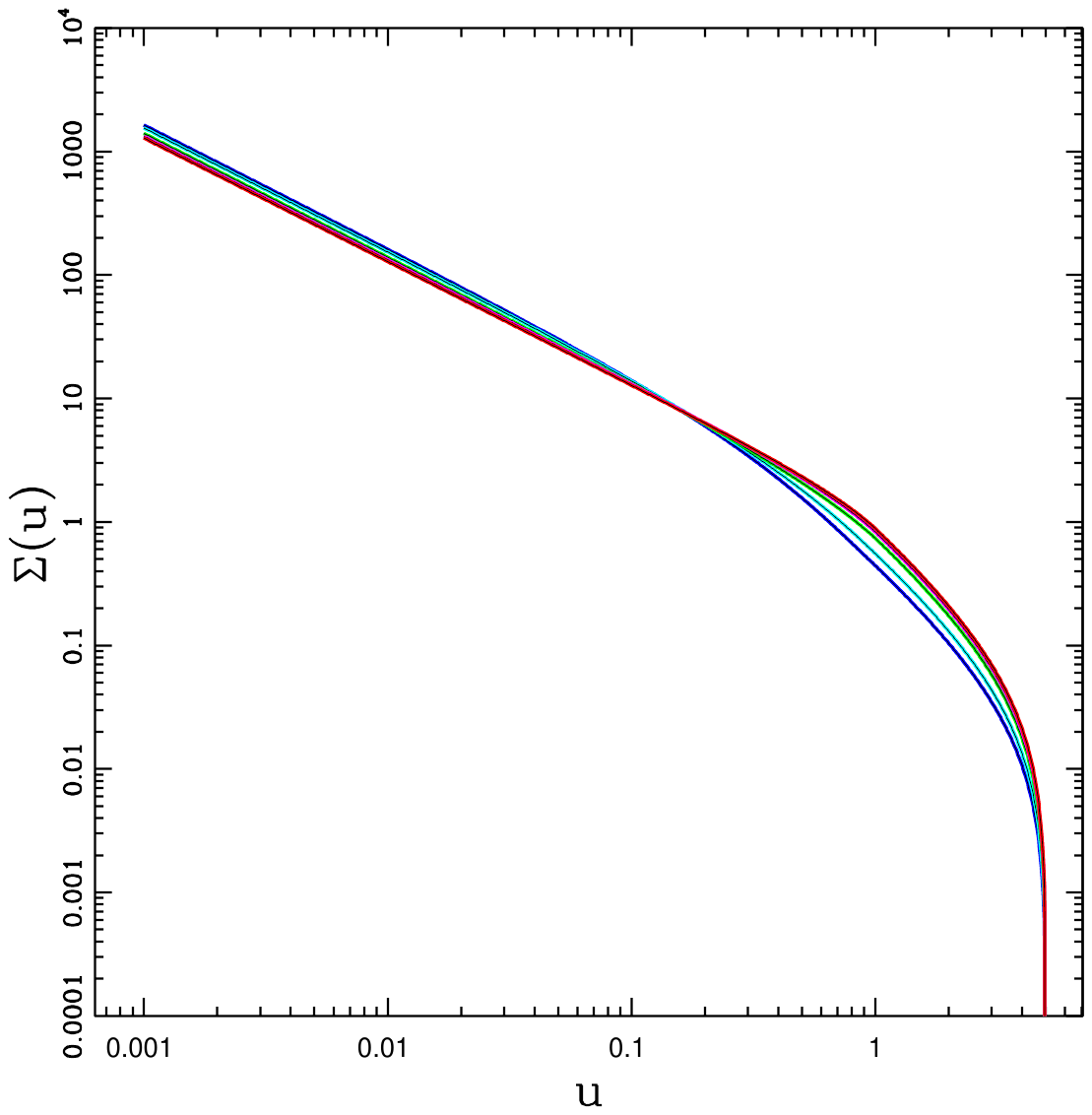}
\vskip-1.0truein
\caption{Surface density distributions for circumplanetary disk being
  fed by varying infall patterns, from polar (top blue curve on the
  left) to isotropic (middle green curve) to equatorial (bottom red
  curve on the left). Here the outer boundary is specified by
    $\uout$ = $3\xi/\lambda^2$ = 5 (so that $\lambda\sim1/2$).} 
\label{fig:sigma}
\end{figure}

\begin{figure}
\centering
\includegraphics[width=0.80\linewidth]{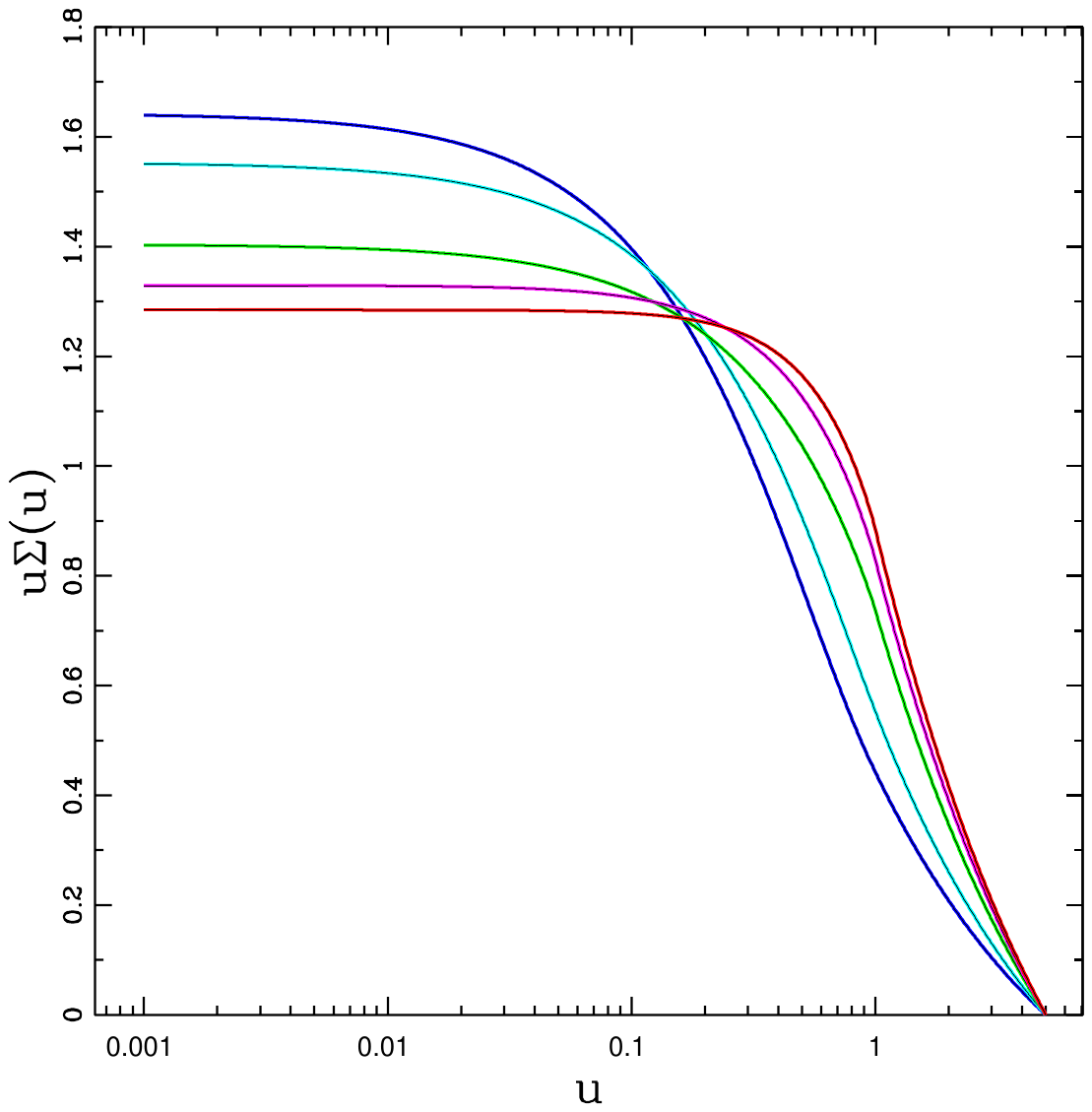}
\vskip-1.0truein
\caption{Surface density distributions for circumplanetary disk scaled
  by one power of the radial coordinate. The various curves correspond
  to different infall patterns, from polar (top blue curve on the
  left) to isotropic (middle green curve) to equatorial (bottom red
  curve on the left). Here the outer boundary is specified by
    $\uout$ = $3\xi/\lambda^2$ = 5 (so that $\lambda\sim1/2$).} 
\label{fig:sigscale}
\end{figure}

The surface density profiles are depicted in Figures \ref{fig:sigma}
and \ref{fig:sigscale} for the five flow geometries under
consideration. Figure \ref{fig:sigma} shows the surface density
itself, in dimensionless form, i.e., the functions $u^{-1} s_k(u)$ for
$k$ = 1 -- 5. These forms are valid for any choice of the mass infall
rate $\mdotin$, viscosity scale $\nu_C$ (and $\alpha$), and angular
momentum bias $\lambda$ (which sets the centrifugal radius $R_C$).  To
leading order, the surface density distributions show simple power-law
behavor $\Sigma\propto1/u$ in the inner regions, with a cutoff at the
outer disk edge (here at $\uout$ = 5). The departures from a power-law
form are necessary to reach a steady-state configuration for the given
source functions, and to satisfy the outer boundary condition.  Figure
\ref{fig:sigscale} shows the corresponding profiles scaled by one
power of radius, i.e., the functions $s_k(u)$. These scaled profiles
reach a constant value in the inner regions ($u\to0$), where the value
is proprotional to the accretion efficiency (see equation
[\ref{efficiency}]).  In both figures, the colors depict the results
for the different input geometries, varying from polar (top blue
curve) to isotropic (middle green curve) to equatorial (bottom red
curve).

\begin{figure}
\centering
\includegraphics[width=0.80\linewidth]{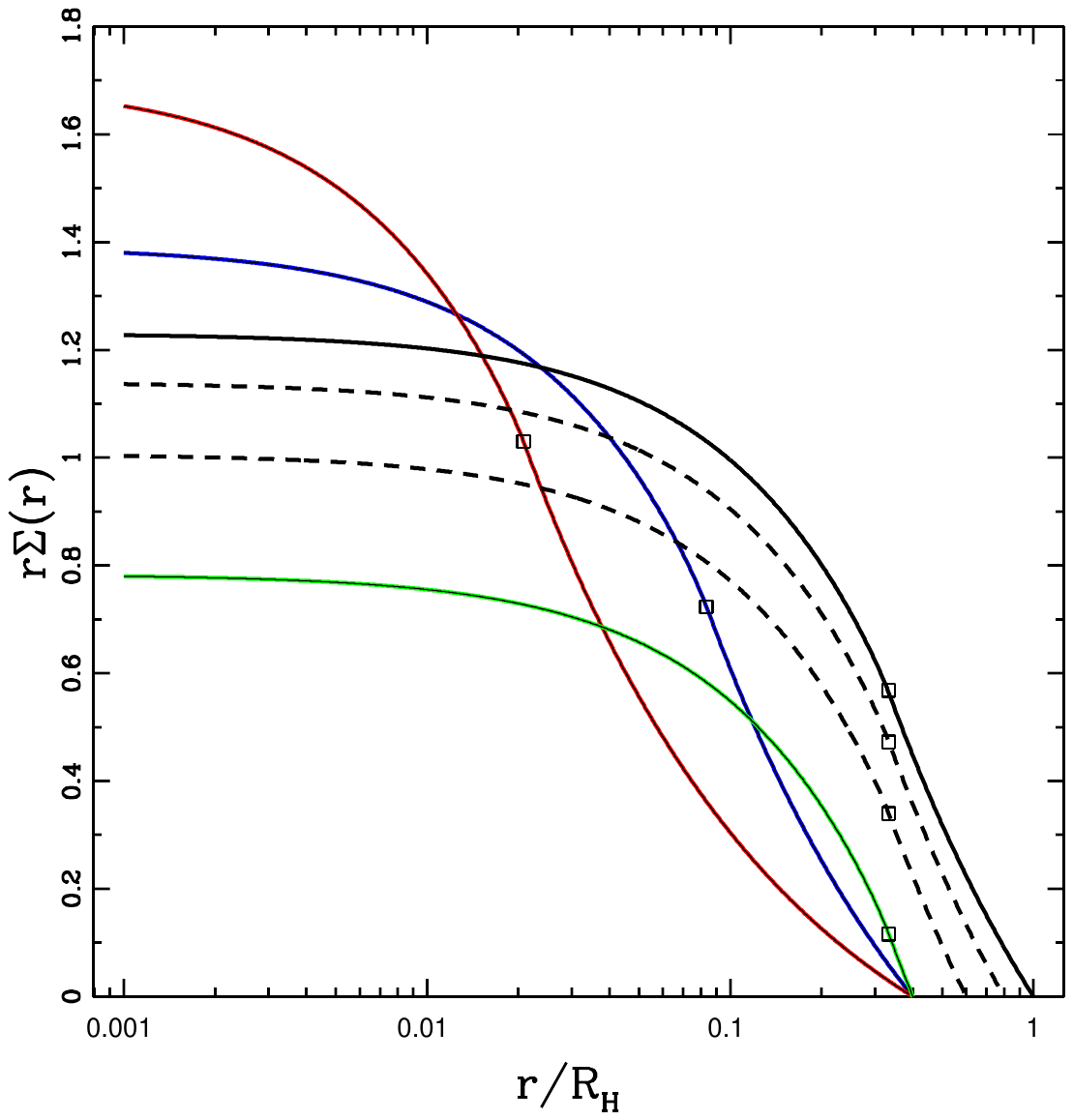}
\vskip-1.0truein
\caption{Surface density distributions for the circumplanetary
  disk for varying values of angular momentum bias factor $\lambda$
  and outer disk boundary parameter $\xi$. The three colored curves
  show the surface density as a function of $r/R_H$ with a fixed outer
  boundary given by $\xi$ =2/5 and angular momentum bias factors
  $\lambda$ = 1 (green), 1/2 (blue), and 1/4 (red). The solid black
  curves shows the surface density for $\lambda$ = 1 and $\xi$ = 1.
  The dashed black curves correspond to $\lambda$ = 1 and $\xi$ = 0.8
  (upper) and 0.6 (lower). The square symbols mark the location of the
  centrifugal radius $R_C/R_H=\lambda^2/3$. }
\label{fig:sigvary}
\end{figure}

Figure \ref{fig:sigvary} shows the effects of varying the angular 
momentum bias factor $\lambda$ and the location of the disk outer
boundary as determined by $\xi$ (recall that $\rout=\xi R_H$). The
scaled surface denisty $r\Sigma$ is plotted versus $r/R_H$, where the
radius is shown on a logarithmic scale. For simplicity, surface
density distributions are shown for the case of steady-state evolution
and an isotropic infall pattern.  In the figure, the three colored
curves depict the surface density distributions for angular momentum
bias values $\lambda$ = 1 (green), 1/2 (blue), and 1/4 (red). For all
three profiles, the outer boundary is held constant at
$\rout=2R_H/5$. As the angular momentum bias factor decreases, the
surface density increases at small radii, and the profile decreases
more rapidly with $r$, thereby leading to an effectively steeper
density distribution.  In order to illustrate how the surface density
profiles depend on the outer boundary condition, the black curves show
profiles with fixed angular momentum bias $\lambda$ = 1, and varying
values of $\xi$ = 1 (upper solid curve), 0.8 (middle dashed curve),
and 0.6 (lower dashed curve). Note that the green curve (with
$\lambda$ = 1 and $\xi$ = 0.4) continues the sequence. The square
symbols show the locations of the centrifugal radius $R_C$ for each
case (where $R_C/R_H=\lambda^2/3$). As the location of the outer
boundary moves inward, the scaled surface density profile decreases by
an overall factor, but retains essentially the same shape. 

Although the surface density profiles found here (equations
[\ref{surf1} -- \ref{surf5}]) are expressed in terms of elementary
functions, their forms remain somewhat complicated. For some
applications, even simpler approximations are useful. Toward this end,
Appendix \ref{sec:fitting} presents a straightforward approximation
scheme that captures the basic properties of these solutions
(including how the profiles vary with infall geometry). 

\subsection{Circumplanetary Disk Mass}
\label{sec:diskmass} 

The mass of the circumplanetary disk is given by integrating
the surface density over the disk area, i.e., 
\be
M_{\rm disk} = \int_0^{R_H} 2\pi r dr \Sigma (r) 
 = {{\dot M} R_C^2 \over 3 \nu_C} \int_0^{\uout} 
s_k(u) du \equiv {{\do t M} R_C^2 \over 3 \nu_C} I\,,
\ee
where the final equality defines the dimensionless integral $I$.
We can integrate the profiles from the previous subsection to
obtain
\be
I = {4\over5}\left(1+\sqrt{\uout}\right) - {3\pi\over8} \,,\qquad
\sqrt{\uout} - {3\over5} \,,\qquad
{4\over3}\left(1+\sqrt{\uout}\right) - {3\pi\over4} \,,
\ee
$$
{3\over2}\left(1+\sqrt{\uout}\right) - {128\over15\pi} \,,\qquad
{\rm and} \qquad 
{8\over5}\left(1+\sqrt{\uout}\right) - {15\pi\over16} \,.
$$
The characteristic mass scale for the disks is given by the
leading coefficient, which can be evaluated for typical
parameters to obtain
\be
M_\ast \equiv {{\dot M} R_C^2 \over 3 \nu_C} \approx
3.9\times10^{-4}\,\mjup\,
\left({\alpha\over 10^{-3}}\right)^{-1} 
\left({ {\dot M} \over 1 \mjup {\rm Myr}^{-1}} \right)
\left({M\over1\mjup} \right)^{2/3} \,, 
\ee
where we have assumed that the planet resides at $a=5$ AU, the
stellar mass $M_\ast=1M_\odot$, and the angular momentum bias
$\lambda=\sqrt{6}/6$. For completeness, the corresponding
surface density for a disk in steady-state accretion is
determined by the leading coefficient
\be
\Sigma_\ast \equiv { {\dot M} \over 6\pi\nu_C} \approx 1400 \,
{\rm g} \, {\rm cm}^{-2} \left({\alpha\over 10^{-3}}\right)^{-1} 
\left({ {\dot M} \over 1 \mjup {\rm Myr}^{-1}} \right)\,. 
\ee
It is instructive to compare these values with the properties of
circumstellar disks that provide the planet forming environment.
Circumplanetary disks are expected to have much lower masses in
steady-state, $M_{\rm d}\sim0.001\mjup$ compared to $\sim50\mjup$ for
the Minimum Mass Solar Nebula (MMSN; e.g., \citealt{hayashi1981}).
On the other hand, the surface densities are comparable: The MMSN
has $\Sigma\sim1000$ g cm$^{-2}$ at 1 AU.

\subsection{Disk Temperature Distributions}
\label{sec:temperature}

The above analysis determines the mass accretion rate through the disk
and the corresponding surfaces density profiles.  These solutions, in
turn, determine the temperature distribution of the circumplanetary
disk due to accretion. The energy dissipated per unit time per unit
energy is given by
\be
D(r) = \nu \Sigma \left( r {d\Omega\over dr}\right)^2  =
{9\over4} \nu \Sigma {GM \over r^3} \,, 
\ee
where the final equality holds for a Keplerian rotation curve
(as expected here). If the disk is sufficiently optically thick,
it develops a photosphere with a well defined temperature, and
radiates like a blackbody to leading order.  Using the surface
density distributions of this paper, specifically the form of
equation (\ref{gensigma}), the disk temperature is given by
\be
\sigma T^4 = {3\over8\pi}{G M \mdotin \over r^3} s_k(u) \,,
\label{disktemp} 
\ee
where $u=r/R_C$. As a result, the temperature departs from the
familiar form $T \sim r^{-3/4}$ by a factor of $[s_k(u)]^{1/4}$.
This correction factor approaches a constant of order unity in the
inner limit $u\ll1$, and tapers off so that the temperature vanishes
in the limit $u\to\uout$ (where $s_k\to0$). Note that the quantity
$\mdotin$ is the total rate at which the planet/disk system gathers
fall from the infalling envelope. Only a fraction $\gamma_{\rm in}$ of
this mass is accreted by the planet, with the remainder transfered
outward to conserve angular momentum. The required efficiency factor
is included in the specification of the functions $s_k(u)$.

Note that the temperature given by equation (\ref{disktemp})
represents the contribution provided by disk accretion only. In
addition, the material falling onto the disk will dissipate some of
its energy in a shock on the disk surface \citep{adams1986,adams2022}
and will dissipate additional energy as it adjusts to the the
pre-existing Keplerian rotation profile (\citealt{cassen1981}; Section
\ref{sec:adjust}). Radiation from both the central planet and the
background circumstellar disk provide additional contributions to the
heating of the surface of the circumplanetary disk. As a result, the
effective surface temperature of the disk will be hotter than that
given by equation (\ref{disktemp}).

\section{Conclusion}
\label{sec:conclude}

This paper considers the formation of circumplanetary disks that form
alongside their host planets during the late stages of giant planet
formation. Due to conservation of angular momentum, most of the mass
that eventually accretes onto the forming planet is processed through
the disk.  These disk structures are crucial for transporting angular
momentum outward, so that mass can be be transferred onto the planet.

\subsection{Summary of Results} 
\label{sec:summary} 

This paper generalizes previous calculations in several ways: [a] We
include the angular momentum bias $\lambda$, which accounts for the
fact that the incoming material -- at the outer boundary of the Hill
sphere -- does not necessarily have the same rotation rate as the mean
motion of the planet. The bias factor is expected to lie in range
$\lambda\approx1/3-1/2$ during the late stages of formation (see
\citealt{ward2010} and references therein).  [b] We generalize the
orbit solutions that describe the trajectories of incoming material
(see Section \ref{sec:orbit} and \citealt{mendoza2009}) starting at
the Hill sphere and eventually falling onto the surface of the
circumplanetary disk.  [c] We modify the outer boundary of the
circumplanetary disk. Instead of extending all the way out to the Hill
radius, the disk will be effectively truncated at a smaller distance
$\rout$ due to orbit crossing.  As a result, we enforce the outer
boundary conditions at $\xi=\rout/R_H \approx0.41$ \citep{martin2011}.
[d] Finally, we account for the redistribution of incoming material
after it falls onto the circumplanetary disk. The newly added material
does not, in general, have the angular momentum appropriate for a
Keplerian orbit around the planet at the location where it joins the
disk \citep{cassen1981,cassen1983}.  As a result, new material mixes
with pre-existing disk gas and dissipates energy while conserving
angular momentum. This readjustment changes the form of the source
term for material being added to the disk (equation [\ref{source}]).
The effects of angular momentum adjustment on the surface
densities are outlined in Appendix \ref{sec:adcompare}. 

Including the aforementioned generalizations, we have constructed
analytic solutions (Section \ref{sec:steadystate}) for the disk
surface density distribution $\Sigma(r)$ for a range of infall
geometries (as characterized by the geometrical functions $f(u)$; see
equation [\ref{fufun}]). The resulting surface density profiles have
the approximate form $\nu\Sigma\to$ {\sl constant} in the inner limit
$r\to0$, and smoothly decrease beyond the centrifugal barrier so that
$\Sigma\to0$ at the outer disk edge (see Figures \ref{fig:sigma} and
\ref{fig:sigscale}).  For expected parameters, the viscosity
$\nu\propto{r}$ so that $\Sigma\sim1/r$ over most of the disk within
the centrifugal radius. Note that the functional forms for the surface
density distributions are relatively insensitive to the angular
distribution of the incoming material (set by $f[u]$) and have the
same general form as found previously.

Although the general form for the surface density profiles $\Sigma(u)$
are robust, the generalizations of this paper lead to important
corrections (see Figure \ref{fig:sigvary}).  The inclusion of
the angular momentum bias reduces the size of the centrifugal barrier
$R_C$ by a factor of $\lambda^2\sim1/9-1/4$ (for expected values
$\lambda=1/3-1/2$). For the benchmark case of a 1 $\mjup$ planet
forming at $a$ = 5 AU, we find $R_C\approx3\times10^{11}$ cm $\approx$
30 $R_P$. In terms of the present-day Jovian satellite system, this
estimate for the centrifugal barrier lies outside the orbit of
Callisto and inside the orbit of the irregular moon Themisto. In
addition to resulting in smaller disks, the smaller values for $R_C$
lead to larger values for the column density of the infalling envelope
surrounding the planet. The column density scales as $N_{\rm col}
\propto R_C^{-1/2}$ $\propto \lambda^{-1}$
\citep{adams1986,adams2022}.

In general, accretion disks have inward flow in the inner limit and
outward flow at their outer boundary. The radius $\uzero$ where the
mass accretion rate vanishes (so that ${\dot M}$ changes sign) plays
an important role in determining disk structure, and has important
implications for the formation of Jovian moons
\citep{canup2002,batygin2020}.  Although the geometric distribution of
incoming material has only a modest effect on the general form of the
suface density, the values of $\uzero$ depend on the choice of
$f(u)$. For the five cases considered here, the we find $\uzero$ =
0.350, 0.482, 0.751, 0.800, and 0.832 for $k=1-5$ (polar to isotropic
to equatorial concentrations). The physical radius where the accretion
flow changes sign is given by $R_0=u_0R_C=u_0\lambda^2R_H/3$.

In these systems, the circumplanetary disk initially accretes the
majority of the mass and angular momentum impinging upon the central
object.  Since the planet itself carries little angular momentum,
relative to the total incoming amount (even if the planet spins at
breakup), the disk must transfer essentially all of the angular
momentum outward to the outer disk boundary, where it joins the
reservoir of the background circumstellar disk. Conservation of
angular momentum necessarily results in some loss of mass, so that the
accretion process cannot be fully efficient. The accretion efficiency
is determined by the fraction of the infalling material that accretes
onto the central planet (see Section \ref{sec:steadystate}), and can
be written in the form  
\be
\gamma_{\rm in} = 1 - {{\cal F}_k \lambda \over \sqrt{3\xi}}\,,
\label{efficiency} 
\ee
where ${\cal F}_k$ is the dimensionless angular momentum for a given
infall geometry (labeled by the index $k$; see equation
[\ref{jfive}]), $\lambda$ is the angular momentum bias, and $\xi$
defines the outer boundary of the disk $\rout$ = $\xi R_H$. For
expected values of the parameters, the efficiency lies in the range
$\gamma_{\rm in}\approx$ 0.64 -- 0.82.  Since $\lambda^2<\xi$,
inclusion of both the angular momentum bias and corrections
to the outer disk boundary result in a {\it higher} efficiency
compared to previous treatments.\footnote{Keep in mind that the
efficiency $\gamma_{\rm in}$ is the fraction of the incoming material
that strikes the disk and is then accreted onto the planet. A second
efficiency factor is also present: Only a fraction of the material
that enters the Hill sphere stays within the vincinity of the planet
and falls onto the disk. In this treatment, the incoming mass
accretion rate $\mdotin$ includes only the material that stays within
the Hill sphere.}

\subsection{Discussion}
\label{sec:discuss}

The results of this paper define length scales that characterize the
properties and drive the evolution of circumplanetary disks. In
addition to the planetary radius $R_P$, previous work has defined the
magnetic truncation radius $R_X$ \citep{ghosh1978,blandford1982}, the
centrifugal radius \citep{ulrich1976,quillen1998}, and invoked the
Hill radius $R_H$ as the boundary for the planetary sphere of
influence. In addition, this work sets the outer disk boundary $\rout$
at a fraction of the Hill radius, determines the radii $R_0$ where the
mass accretion flow changes sign, and makes the distinction between
the centrifugal radius and the initial disk radius $R_{\rm d}$ (which
depends on the starting radial velocity -- see equation
[\ref{diskrad}]). These length scales obey the ordering 
\be
R_P < R_X < R_0 < R_{\rm d} \lta R_C < \rout < R_H \,.
\label{ordering} 
\ee 

This paper has generalized the approach used in previous work
regarding the starting conditions at the Hill sphere, the outer disk
boundary, and manner in which incoming material enters the disk. Like
any analytical treatment, this paper makes approximations, including
that of azimuthal symmetry, neglect of magnetic fields, and the use of
ballistic trajectories (see \citealt{adams2022} for a more detailed
discussion of their validity). On this latter issue, we note that the
infalling envelope must be able to cool sufficiently in order for the
circumplanetary disk to form. This cooling requirement implies an
upper limit to the mass infall rate $\mdotin$. More specifically,
successful disk formation requires the cooling time of the envelope to
be shorter than the free-fall time -- analogous to classic arguments
regarding opacity limited fragmentation (see \citealt{rees1976}). This
constraint, derived in Appendix \ref{sec:cooling}, implies the upper
bound $\mdotin\lta70$ $\mjup$ Myr$^{-1}$. The other key assumption is
that the circumplanetary disk can reach steady-state, which requires
the viscous evolution time $t_\nu = R_C^2/\nu$ to be shorter than the
infall time scale $t_{\rm in}=M/\mdotin$, a constraint that is readily
met (see Section \ref{sec:steadystate}).

The surface density profiles found in this paper apply to late stages
of the core accretion paradigm for planet formation, when the growing
planet accumulates the majority of its mass. During this time, the
Hill radius is smaller than the Bondi radius, but the planet remains
(at least mostly) embedded within its parental circumstellar disk. At
still later times, however, the Hill radius can exceed the scale
height of the circumstellar disk and the planet can clear a gap in the
nebula. Both of these circumstances act to reduce the amount of gas
flowing into the Hill sphere, eventually leading to a decrease in
$\mdotin$, so that we expect most of the planetary mass to be accreted
beforehand. Nonetheless, these final stages are important for
determining the final mass of the planet and for the issues related to
satellite formation. As such, these end stages should be considered in
future work.

\bigskip 

We are grateful to A. Taylor for useful discussions. This work was
supported by the University of Michigan, the California Institute of
Technology, the Leinweber Center for Theoretical Physics, and by the
David the Lucile Packard Foundation.

\appendix
\section{Cooling Constraint}
\label{sec:cooling}

As outlined in Section \ref{sec:intro}, numerical simulations show
that the formation of circumplanetary disks requires sufficient
cooling. Recent multi-fluid simulations \citep{krapp2024} indicate
that the gas must cool on a timescale at least ten times shorter than
the orbital timescale in order to retain high angular momentum and
form a disk. If the cooling time is longer (cooling is inefficient),
the outcome is an isentropic, convective envelope that extends through
much of the Hill sphere and rotates far below the Keplerian speed. In
essence, rapid cooling allows the inflow to become rotationally
supported, whereas slow cooling leads to adiabatic compression that
thermally `inflates' the gas, wiping out any nascent disk.  This
thermodynamic criterion thus sets a high bar for circumplanetary disk
formation. To understand the origin of this criterion, this Appendix
finds the conditions required for the cooling time of the infalling
envelope to be shorter than the free-fall time.  This condition,
$t_{\rm cool}<t_{\rm ff}$, is required for disk formation and can
also be used to place a limit on the infall rate.


The cooling time for the envelope can be written in the form  
\be
\tcool = { \uenv \over 4\pi r^2 \sigma T^4} \,\taueff \,,
\ee
where $\uenv$ is the thermal energy of the infalling envelope, i.e.,
the energy that must be radiated away in order for the flow to
continue. The scale $r$ is some characteristiec radius. The effective
optical depth $\taueff$ can be written in terms of the actual optical
depth $\tau$ \citep{rees1976} according to the relation
\be
\taueff = {1\over2} \left( \tau + {1\over\tau} \right)\,. 
\ee
The free fall time can be written in the form 
\be
t_{\rm ff} = {\sqrt{2}\over3} \left({GM \over R^3}\right)^{-1/2} \,,
\label{freefall} 
\ee
where $R$ is radius where effective cooling is enforced. In order for
disk formation to take place, this radius must be comparable to the
disk size so that $R=R_C$. At this radius, the free-fall time from
equation (\ref{freefall}) is about an order of magnitude shorter than
the orbit time of the forming planet (so that the requirement
$t_{\rm cool}<t_{\rm ff}(R_C)$ is consistent with that advocated by
\citealt{krapp2024} based on numerical simulations). To evaluate the
thermal energy of the infalling envelope, we find the energy that
would be present if the gas heats up adiabatically according to
$P\propto\rho^{5/3}$ so that $v_s^2\propto$ $T\propto\rho^{2/3}$.
We then integrate over the envelope using the density distribution
for the infall to find 
\be
\uenv = \int_0^{R_H} 4\pi r^2 dr \rho v_{sH}^2 (\rho/\rho_H)^{2/3} =
8\pi C R_H^{3/2} v_{s}^2 =
{\sqrt{2} \mdotin R_H^{3/2} \over \sqrt{GM}} v_{s}^2(R_H) \,, 
\ee
where the sound speed $v_s$ is evaluated at the Hill radius. After
combining these results, the cooling constraint takes the form
\be
4\pi r^2 \sigma T^4 > \mdotin 3 v_s^2 
\left({R_H\over R_C} \right)^{3/2} \taueff \,. 
\ee
The left-hand-side of the equation is determined by the envelope
luminosity, which must be a fraction of the total system luminosity,
so that 
\be
4\pi r^2 \sigma T^4 = \beta {GM \mdotin \over R_P}
\left[1 - \exp(-\tau_\ast)\right] \,.
\label{luminosity} 
\ee
This expression assumes that the total luminosity is given by the
incoming material falling to the planetary surface. The envelope must
radiate a fraction of this total power as determined by the optical
depth $\tau_\ast$ of the envelope to the radiation from the central
source. Note that this optical depth $\tau_\ast$ is generally larger
than the optical depth $\tau$ of the envelope to its internally
emitted radiation. The dimensionless parameter $\beta\lta1$, where the
maximum value corresponds to maximally efficient accretion, with all
of the gas falling to the planetary surface and no energy stored in
rotation. Using the expression (\ref{luminosity}) for luminosity 
in the cooling constraint, we find 
\be
\Lambda \left[1 - {\rm e}^{-\tau_\ast}\right] > \tau + {1\over\tau} \,,
\ee
where the dimensionless parameter $\Lambda$ is defined by 
\be
\Lambda \equiv \beta {2GM \over 3v_s^2 R_P}
\left({R_C\over R_H} \right)^{3/2} \approx 20 \beta
\left({T_H\over134\,{\rm K}}\right)^{-1}
\left(\sqrt{6}\lambda\right)^3 \,. 
\ee
For much of the parameter space of interest, the optical depth of the
central source radiation is large enough that we can ignore the
decaying exponential $\exp[-\tau_\ast]$, so the constraint simplifies
to the form 
\be
\tau \lta \Lambda \approx 20 \qquad {\rm where} \qquad \tau\approx0.3
\left({\mdotin \over 1 \mjup/{\rm Myr}}\right)
\left({M\over1\mjup}\right)^{-2/3}\,. 
\ee
The corresponding constraint on the mass infall rate becomes 
\be
\mdotin \lta 70 \,\mjup \,{\rm Myr}^{-1}\,
\left({M\over1\mjup}\right)^{2/3}\,. 
\ee
Note that we have some flexibility in choosing numbers to evaluate the
constraint, so a more conservative limit would be $\mdotin\lta100$
$\mjup$/Myr. In any case, the infall rate cannot become arbitarily
large and still allow for the envelope to cool, form a disk, and
effectively transfer material onto the planet.

\section{Simple Fitting Functions}
\label{sec:fitting}

Although this paper has successfully obtained analytic forms for the
surface density profiles of circumplanetary disks, the resulting forms
are somewhat complicated (see Section \ref{sec:steadystate}). On one
hand, these complicated functions are necessary for the surface
density profiles to follow the angular momentum budget for a given
source function. On the other hand, simpler but more approximate
forms are useful in some applications. The surface density profiles
for the five geometries of interest can be written in the form
\be
\Sigma = \Sigma_0 \, u^{-1} \exp[-u/2u_0] \,, 
\ee
where $u=r/R_C$, $u_0$ is the location in the disk where the mass
accretion rate through the disk changes sign, and the leading 
constant is given by  
\be
\Sigma_0 = {\mdotin \over 3\pi \nu_C}
\left[1 - {{\cal F}_k \over \sqrt{\uout}} \right] \,, 
\ee
where the angular momentum factors ${\cal F}_k$ for the five cases
of interest are given by equation (\ref{jfive}) and where
$\uout=\rout/R_C=3\xi/\lambda^2$. Note that all of the constants in
the expression are determined by the properties of the full analytic
solutions. With these specifications, by construction, the surface
density has the (exact) correct form in the limit $u\to0$ and the
corresponding mass accretion rate changes sign at the proper location.
The largest departure of the fitting formula from the exact solutions
occurs near the outer disk edge where $\Sigma(\uout)\to0$. The
resulting fits are thus approximate. If we define the root-mean-square
errors according to 
\be
{\cal E}_k \equiv \left[
\int_0^{\uout} (s_k - s_{k{\rm fit}})^2 du \right]^{1/2} 
\left[\int_0^{\uout} s_k du \right]^{-1}, 
\ee
then the errors ${\cal E}_k$ $\approx$ 0.11, 0.058, 0.030, 0.070, and
0.095 for the five cases shown in Figure \ref{fig:sigscale}. For the
infall profiles concentrated along the equator ($k=4,5$), the fitting
functions perform poorly for radii beyond the centrifugal barrier
($u>1$); the errors are not overly large because the profiles decrease
rapidly in this regime. One can readily find more accurate fitting
formulae -- but at the expense of needing more complicated
expressions.

As one example of the utility of these fitting functions, the
total disk accretion luminosity can be written in the form
\be
L_{\rm ac} = {3GM\mdotin\over2 R_X} \int_1^\infty
{d\xi\over\xi^2} \exp[- q \xi] =
{3GM\mdotin\over2 R_X} {\rm E}_2(q) \,, 
\ee
where $q=R_X/(2u_0R_C)$ and $E_2$ is the exponential integral
\citep{abrasteg}. Recall that the disk is truncated on the inside
at $R_X$ due to magnetic fields from the planet.

\section{Effect of Angular Momentum Adjustment on the Surface Density}
\label{sec:adcompare}

This Appendix compares the surface density solutions of this paper
with those found earlier without angular momentum adjustment.  In
addition to angular momentum bias and modified outer boundary
conditions, this paper explicitly includes the redistribution of
incoming material as it impacts the circumplanetary disk. The original
source term from equation (\ref{source}) specifies the rate at which
the surface of the disk gains mass. Since the incoming material does
not have the proper angular momentum, that appropriate for a Keplerian
rotation curve, it adjusts its position accordingly (see Section
\ref{sec:adjust}). Due to this adjustment, the effective source term
is modified, as shown by the first term in the surface density
evolution equation (\ref{steady}). In order to isolate the effects of
angular momentum adjustment on the resulting surface density profiles,
we determine the analogs of the functions $s_k(u)$ from Section
\ref{sec:disksummary}. Here we use equation (\ref{source}) as the
source term, without angular momentum adjustment, so that the disk 
evolution equation simplifies to the form
\be
\source(u) + {3 \over R_C^2}{1\over u} {\partial\over\partial u}
\left( \sqrt{u} {\partial\over\partial u}
\left[ \sqrt{u} \nu \Sigma\right] \right) = 0 \,,
\ee
where the source term includes the geometric function $f_k(u)$ for the
different infall geometries. We use the same angular momentum bias and
disk outer boundary as before, so that $\uout=5$. The resulting forms
for the surface density profiles are those given by equations (41--46)
from \cite{taylor2024}, except that these earlier solutions were 
expressed in terms of arbitrary constants $K_1$ and $K_2$ (see also
\citealt{adams2022}). Here we specify the constants to meet the
aforementioned boundary conditions and to be consistent with the
solutions from the present paper.

\begin{figure}
\centering
\includegraphics[width=0.80\linewidth]{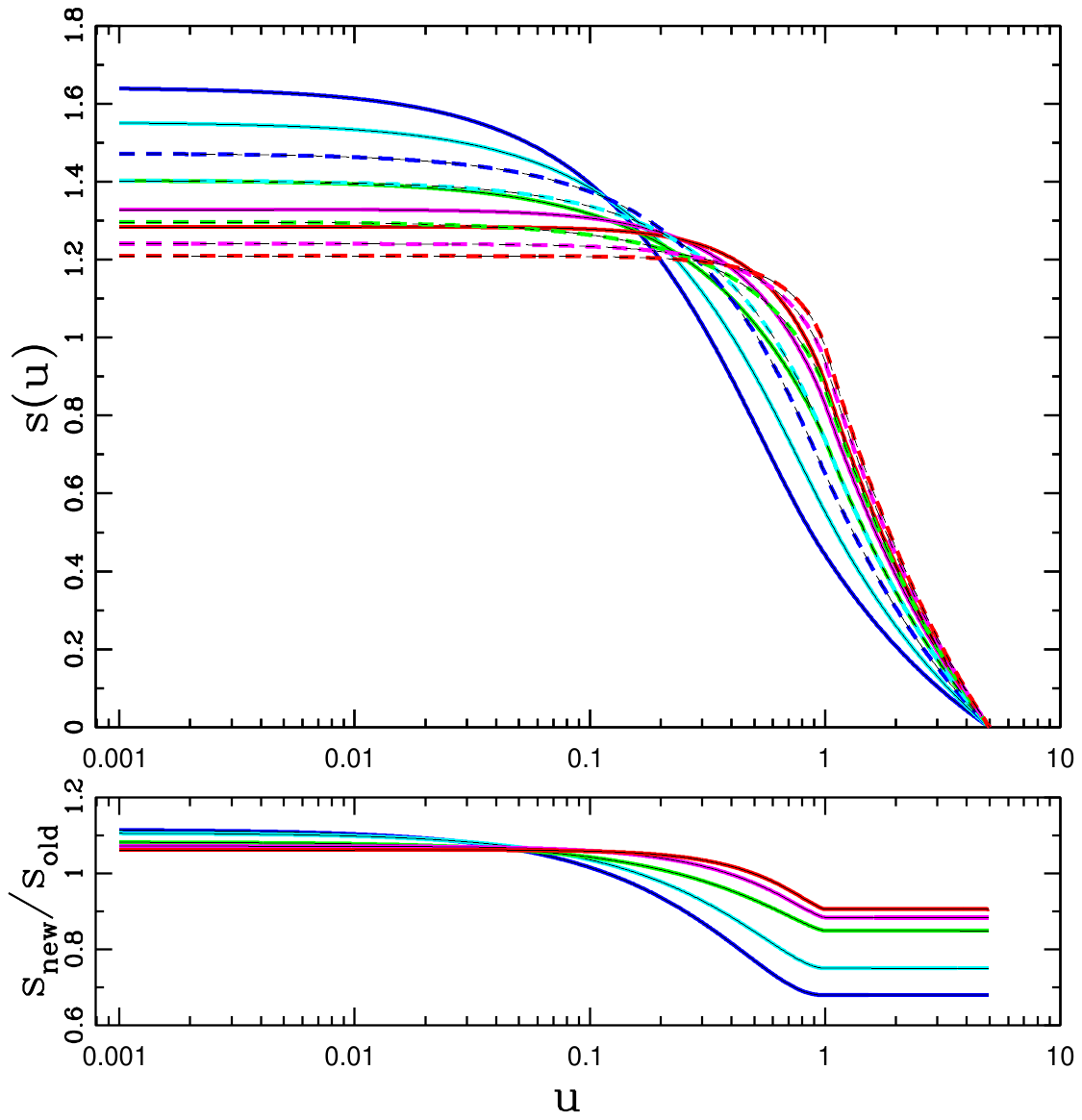}
\vskip-1.0truein
\caption{Surface density distributions for the circumplanetary
  disks with and without angular momentum adjustment (see text). In
  the upper panel, solid curves show the solutions with angular
  momentum adjustment, for polar flow ($k=1$, solid blue) to isotropic
  ($k=3$, solid green), to equatorial ($k=5$, solid red). The dashed
  curves show the corresponding sufface density profiles calculated in
  the absence of angular momentum adjustment. The lower panel shows
  the ratio of the profiles with adjustment ($s_{\rm new}$) to those
  without ($s_{\rm old}$), where the index runs from $k=1$ (upper blue
  on the left) to $k=5$ (lower red on the left).}
\label{fig:angmom}
\end{figure}

The resulting surface density profiles, both with and without angular
momentum adjustment, are presented in Figure \ref{fig:angmom}. In the
upper panel, the solid curves show the surface density profiles
$s_k(u)$ from this paper (the same as in Figure \ref{fig:sigscale}).
The five solutions for the different infall geometries are shown in
blue ($k=1$), cyan ($k=2$), green ($k=3$), magenta ($k=4$), and red
($k=5$), from top to bottom on the left side of the plot. The
corresponding surface density profiles, calculated without angular
momentum adjustment, are shown as the dashed curves (with the same
ordering and colors). The bottom panel of Figure \ref{fig:angmom}
shows the ratio of the surface density profiles for the five infall
geometries. These results show that the inclusion of angular momentum
adjustment makes the surface density relatively higher at small radii
and correspondingly smaller at larger radii. In other words, the
surface density profiles become steeper. This trend is expected, as
the adjustment process (Section \ref{sec:adjust}) acts to move
incoming material inward. The magnitude of the effect is modest,
however, as shown in the lower panel. Over most of the radial range
of interest, the difference is of order 10 -- 20\%.  

\end{document}